\documentclass[12pt,preprint]{aastex}
\newcommand{\kms}{\relax \ifmmode {\,\mbox{km\,s}}^{-1}\else \,\mbox{km\,s}$^{-1}$\fi}
\newcommand{\ergs}{\relax \ifmmode {\,\mbox{erg\,s}}^{\mbox{-1}}\else \,\mbox{erg\,s}$^{\mbox {-1}}$\fi}
\newcommand{\ha}{\relax \ifmmode {\mbox H}\alpha\else H$\alpha$\fi}
\newcommand{\hb}{\relax \ifmmode {\mbox H}\beta\else H$\beta$\fi}
\newcommand{\hi}{\relax \ifmmode {\mbox H\,{\sc i}}\else H\,{\sc i}\fi}
\newcommand{\hei}{\relax \ifmmode {\mbox He\,{\sc i}}\else He\,{\sc i}\fi}
\newcommand{\heii}{\relax \ifmmode {\mbox He\,{\sc ii}}\else He\,{\sc ii}\fi}
\newcommand{\hii}{\relax \ifmmode {\mbox H\,{\sc ii}}\else H\,{\sc ii}\fi}
\newcommand{\oiii}{\relax \ifmmode {\mbox O\,{\sc iii}}\else O\,{\sc iii}\fi}
\newcommand{\oii}{\relax \ifmmode {\mbox O\,{\sc ii}}\else O\,{\sc ii}\fi}
\newcommand{\oi}{\relax \ifmmode {\mbox O\,{\sc i}}\else O\,{\sc i}\fi}
\newcommand{\sii}{\relax \ifmmode {\mbox S\,{\sc ii}}\else S\,{\sc ii}\fi}
\newcommand{\siii}{\relax \ifmmode {\mbox S\,{\sc iii}}\else S\,{\sc iii}\fi}
\newcommand{\ariii}{\relax \ifmmode {\mbox Ar\,{\sc iii}}\else Ar\,{\sc iii}\fi}
\newcommand{\ariv}{\relax \ifmmode {\mbox Ar\,{\sc iv}}\else Ar\,{\sc iv}\fi}
\newcommand{\lha}{\relax \ifmmode \mbox {L}_{H\alpha}\else $\mbox{L}_{H\alpha}$\fi}
\newcommand{\ldig}{\relax \ifmmode {\mbox L}_{DIG}\else ${\mbox L}_{DIG}$\fi}
\newcommand{\ls}{\relax \ifmmode {\mbox L}_{ Str}\else ${\mbox L}_{ Str}$\fi}
\newcommand{\eme}{\relax \ifmmode {\,\mbox{pc\,cm}}^{-6}\else \,pc\,cm$^{-6}$\fi}
\newcommand{\Msun}{M$_\odot$}
\newcommand{\Rsun}{R$_\odot$}

\def\arcsec{\hbox{$^{\prime\prime}$}}

\def\cm3{\relax \ifmmode {\,\mbox{cm}}^{-3}\else \,\mbox{cm}$^{-3}$\fi}
\def\cmdos{\relax \ifmmode {\,\mbox{cm}}^{2}\else \,\mbox{cm}$^{2}$\fi}
\shorttitle{Models of NGC~346}
\shortauthors{Rela\~no, Peimbert \& Beckman}

\begin{document}

\title{Photoionization Models of NGC~346}

\author{M\'onica Rela\~no\altaffilmark{1},\altaffilmark{2}}
\author{Manuel Peimbert\altaffilmark{3}}
\and
\author{John Beckman\altaffilmark{1},\altaffilmark{4}}

\altaffiltext{1}{Instituto de Astrof\'{\i}sica de Canarias.}

\altaffiltext{2}{Visiting Astronomer, Instituto de Astronom\'{\i}a, 
Universidad Nacional Aut\'onoma de M\'exico.}

\altaffiltext{3}{Instituto de Astronom\'{\i}a, Universidad Nacional Aut\'onoma de
M\'exico.}

\altaffiltext{4}{Consejo Superior de Investigaciones Cient\'{\i}ficas.}

\begin{abstract}
We present spherically symmetric and plane parallel
photoionization models of NGC~346, an \hii\ region in the Small 
Magellanic Cloud. The models are based on {\sc Cloudy} and on
the observations of Peimbert, Peimbert, \& Ruiz (2000). We find
that approximately 45\% of the H ionization photons escape
from the \hii\ region providing an important ionizing source
for the low density interstellar medium of the SMC. The predicted
$I$(4363)/$I$(5007) value is smaller than that observed,
probably implying that there is an additional source of energy
not taken into account by the models. From the ionization
structure of the best model and the observed line intensities we
determine the abundances of N, Ne, S, Ar, and Fe relative to O.

\end{abstract}


\section{Introduction}

NGC~346 is the most luminous \hii\ region of the SMC, its \ha\
luminosity places it on the boundary between normal and giant
extragalactic \hii\ regions. We have produced photoionization models
of this object for the following reasons: a) to find if NGC~346 is 
density bounded or ionization bounded, b) to determine its chemical composition, and 
c) to study the energy budget of the region. The photoionization models
have been computed with {\sc Cloudy 94} (Ferland 1996; Ferland et al. 1998).

The photoionization models computed in this paper present two advantages
relative to similar models of extragalactic \hii\ regions available in 
the literature: a) there are excellent line
intensity observations available for this \hii\ region 
(Peimbert, et al. 2000, hereinafter Paper I) ; b) the spectral types of
most of the ionizing stars are known which permits us to adopt a 
representative ionizing flux. Most photoionization
models in the literature are based on three assumptions: an IMF, an
analytical fit to the adopted IMF, and a star formation history, 
these three assumptions introduce significant uncertainties in the 
ionizing flux (e.g., Cervi\~no, Luridiana,  \& Castander 2000).

\section{Stellar Content and Photoinization Spectrum}

We estimate directly the photoionization spectrum that will be used
for the models based on the spectral classification of the blue stars
presented by Massey, Parker, \& Garmany (1989).

The ionized region of NGC~346 has a diameter of about 420\arcsec\ and 
from the figures and coordinates presented by Massey et al. (1989) and 
Ye, Turtle, \& Kennicutt (1991) we find that
there are 58 blue stars with a visual apparent magnitude $V<15.6$ inside
the ionized boundaries of NGC~346. The blue stars fainter than $V = 15.6$
are expected to be of spectral type O9 or later and do not contribute
appreciably to the photoionization spectrum. 

Of the 58 stars considered
by us 33 have pre-assigned spectral types and 25 do not. We have counted binary systems
as single stars with the exception of HD 5980 where we consider independently
the two main components (the system might have a third component).

We have assigned
spectral types for the other 25 objects based on their V magnitudes and
assuming that they are on the main sequence, these are the objects in
parentheses presented in Table~1.

To derive the ionizing flux for all the stars in Table~1, with the 
exception of HD 5980, we have adopted the
parameters presented by Vacca, Garmany, \& Shull (1996).

For HD 5980 we have taken the stellar parameters derived by Schweickhardt 
\& Schmutz(1999)
where they have assumed that the luminosity of each component has remained
constant during the 1987-1999 period, but varying the temperature and radius
of component A. For the temperature of component A we have adopted
the value it had at the end of 1990, the time of the observations that
we are trying to match with the models. At this temperature the stellar radius
had the value of 21.4~\Rsun.

The total ionizing flux is $40.08 \times 10^{39}$~erg s$^{-1}$ (see Table~1).
The ionizing fluxes of the objects without spectral types are presented in
parentheses in Table~1; their combined flux amounts to $8.80 \times 10^{39}$~erg s$^{-1}$,
21.9\% of the total value, which is a significant but relatively small
fraction.

We estimate errors of 0.2~dex, 0.2~dex and 0.08~dex for the flux of HD 5980,
the total flux of the stars in parentheses, and the total flux of
all the other stars with known spectral types, respectively; consequently 
we have adopted an error of 0.08~dex for the total ionizing flux.

From the ionizing flux and the effective temperature for each star presented
in Table~1 we have built three different photoionization spectra by adopting : $a$)
blackbodies for all the stars, $b$) a set of stellar models for
$Z_{*}=Z_{\odot}$,
and $c$) a set of stellar models for $Z_{*}=0.2~Z_{\odot}$. These three ionization
spectra were chosen to estimate how sensitive are the properties of the models
to changes in the adopted radiation field and consequently how general or
robust are the results obtained from the models. The low metallicity set is based 
on the gaseous abundances for NGC~346 derived in Paper I and assumes that
20\% of the heavy elements are trapped by dust inside the \hii\ region 
(Esteban et al. 1998).

For the models with $Z_{*}=Z_{\odot}$, 
we chose: Mihalas (1972)  
atmospheres for the O3V star and for 5980A, Kurucz (1991)
atmospheres for the stars of Luminosity Class I, and Schaerer \& de Koter (1997)  
for the rest of the stars in Table~1. 
The models with
$Z_{*}=0.2~Z_{\odot}$ are obtained using the library of stellar spectra built
by Lejeune, Cuisinier, \& Buser (1997); this library is based on three sets of 
atmosphere models and covers a wide range of physical
stellar parameters. The three different ionizing continuum spectra used for the
models are illustrated in Figure~1. The model sets are named as followed;
series of models B, for blackbody spectra, series of models S for stellar
models with $Z_{*}=Z_{\odot}$ and series of models L for stellar spectra from 
Lejeune, Cuisinier, \& Buser (1997). 

Based on the models just described and assuming that all the ionizing
photons are trapped by  
the nebula, the total ionizing flux (that is the same for all the models)
implies an \ha\ flux, $L$(\ha) in erg s$^{-1}$, of 39.216~dex 
for the model with $Z_{*}=Z_{\odot}$, 39.237~dex for the model with
$Z_{*}=0.2~Z_{\odot}$, and 39.25~dex assuming a blackbody 
spectrum. The small differences in the \ha\ flux are due to the difference in
the number of ionizing photons, which for a given ionizing flux does depend on 
the shape of the spectrum.

On the other hand, from the observed flux  of $1.55 \times 10^{-9}$~erg s$^{-1}$cm$^{-2}$
(Kennicutt \& Hodge 1996), an interstellar absorption
correction, $C$(\ha), of 0.10~dex (Paper I), and a distance to the SMC
of 64~kpc (Reid 1999), we obtain a total emitted \ha\ flux of 
$38.98 \pm 0.06$~dex (erg s$^{-1}$). Where the error is due to the combination
of 0.04~dex, 0.02~dex, 0.01~dex errors in the observed flux, the adopted 
distance and $C$(\ha), respectively. 

The difference between the \ha\ flux predicted from the ionizing stellar radiation and 
the \ha\ flux derived from observations amounts to $0.24 \pm 0.10$~dex, $0.26
\pm 0.10$~dex and $0.27 \pm 0.10$~dex for the models with $Z_{*}=Z_{\odot}$,
$Z_{*}=0.2~Z_{\odot}$ and for a blackbody spectrum respectively. These values
imply that about $45 \pm 15\%$ of the ionizing photons escape from NGC~346, and 
that the region is 
density bounded. Moreover, this result is in excellent agreement with the models to
ionize the diffuse interstellar medium proposed by 
Zurita, Rozas, \& Beckman (2000), and with the results of Oey \& Kennicutt (1997),
who find that many \hii\ regions in the Large Magallanic Cloud are density
bounded. This problem is discussed further in section~3.6.

The result
that 45\% of the photons escape is independent of the
electron density distribution, but it does not necessarily 
imply that the \hii\ region is density bounded in all
directions, it is also
posible to have a region which is density bounded in some
directions and ionization bounded in others, these types 
of models require density distributions that are not radially
symmetric, the simplest models that study this effect are
those with a "covering factor", where in some directions all
the photons escape and in others all the photons are trapped.
We have assumed for the spherical models a constant 
density distribution or a density distribution that is
spherically symmetric, therefore by density bounded we
mean that photons escape in all directions. It is beyond the
scope of this paper to model other types of density distributions.

\subsection{Initial Mass Function}

It is important to compare our ionizing flux distribution with
that derived from a Salpeter IMF, for this purpose we will estimate
the IMF of the massive stars responsible of most of the ionization.

In Figure~2 we present the IMF derived from Table~1 under the following
assumptions: a) the main sequence stellar masses have been adopted
as the initial masses, b) the mass of stars that are not in the main 
sequence has been adopted as their main sequence mass, c) for each
star instead of an exact mass value we have adopted a uniform continuous
distribution given by the central value $\pm$ 0.02~dex, d) we have
fixed the lower mass limit of the last bin at 51~\Msun, which implies
that this bin includes only half a star, e) the width of the
bins corresponds to 0.08~dex, f) to derive the slope of
the IMF we only considered the central four bins, we did not include 
the last bin because the IMF is truncated at the high mass end
which is inside this bin. We did not include the bin at the low
mass end due to incompleteness, moreover the fraction of the ionizing
flux due to the stars in this bin is negligible.

The fit to the central four bins defined in Figure~2 is given by

\begin{equation}
    log[N/\Delta ~log~m] = 7.675 - (3.71 \pm 0.4)~log~m
\end{equation}

this equation and point d) of the previous paragraph implies that
the upper mass cutoff is equal to 54.1~\Msun. The slope given by  
equation (1) is considerably steeper than -2.35, 
the Salpeter slope,
and indicates that for NGC~346 the use of the Salpeter slope to
represent the ionizing flux is not adequate. 

There are two other determinations of the IMF for the SMC.
Massey et al. (1989) obtain -2.9 $\pm$ 0.3 in the 9~\Msun\ to
85~\Msun\ mass range for NGC~346 and Humphreys and McElroy (1984) obtain
-3.1 in the 15~\Msun\ to 100~\Msun\ mass range for the SMC. From the data of Humphreys
and McElroy we obtain a slope of -3.7 for the 25~\Msun\ to 100~\Msun\ mass range,
in excellent agreement with the slope derived by us.

Our IMF for NGC~346 implies
that for extragalactic \hii\ regions with an \ha\ flux similar
or smaller than that of NGC~346 it is difficult to determine M$_{\rm up}$,
and that the slope, at least for masses higher than 24~\Msun, might be
steeper than the slope given by a Salpeter IMF. 

To determine the ionizing flux for an extragalactic \hii\ region based on
an analytical function we need to know: a) the slope of the IMF, 
b) the M$_{\rm up}$, c) the fraction of ionizing photons that are trapped by the
\hii\ region, and d) the star formation history of the burst that
produced the \hii\ region. As discussed above these four problems
indicate that for NGC~346 it is considerably better to use the ionizing flux 
determined from the individual stars than to derive the ionizing
flux based on an analytical approximation to the IMF.

\section{Photoionization Models}

All the models discussed in this paper were
computed with {\sc Cloudy}. For all models, with the exception of those with 
enhanced and depleted abundances (models L.5 and L.6), we
adopted the following chemical composition (given in log $N$(X)/$N$(H) + 12): 
He = 10.90, C = 7.39, N = 6.51, O = 8.11, Ne = 7.30, Mg = 5.98, Si = 6.37, 
S = 6.59, Cl = 4.80, Ar = 5.82, and Fe = 5.58. Where the He, N, O, Ne, S 
and Ar are the gaseous abundances derived in Paper I, C/O is the value 
derived by Garnett et al. (1995) for the SMC, and  Mg/O, Si/O and Fe/O 
come from the gaseous abundances derived by Esteban et al. (1998) for 
the Orion nebula.

As mentioned in section 2 we used three different sets of ionizing fluxes:
atmospheres with $Z_{*}=0.2~Z_{\odot}$ (models L), atmospheres with 
$Z_{*}=Z_{\odot}$ (models S),
and blackbody radiation fields (models B). The reason to use these three sets 
of ionizing fluxes is to explore how general are the conclusions based on 
the L models. In Figure~1 we show the three integrated spectra used as the
input ionization for the models.

To compare the observed values with the models computed with {\sc Cloudy}
we used the radial integrations of the models that include the center
and compared them with the average intensities of Regions 3 and 13 which are
close to the center. We took the average of the two
regions to reduce the observational errors and to
obtain a more representative value of a line of sight
towards the center of the object. All the models have an inner radius, $r_{0}$, of $10^{16}$~cm,
with the exception of models L.7 and B.3.

We also present in Tables~2 and 3 the line intensities
of region A (that includes the average intensities of 13 regions) because
the errors are smaller than those of the average intensities of regions
3 and 13 and to show that the differences between the two sets of line intensities
are relatively small. For similar reasons we also present 
$T_{\rm e}$(rad)K and $T_{\rm e}$(vol)K the radial and
the volume averaged temperatures weighted by $N_{\rm e}N(\rm O^{++})$ 
obtained from the models, and $T_{\rm e}$([\oiii])K the volume
averaged temperature obtained from the $I(4363)/I(5007)$ ratio. To compare the 
line intensities with those of regions 3 and 13, all the line intensities
have been radial averaged. The main reason for selecting the line ratios 
presented in Tables~2,~3 and 4
is that they are very sensitive to changes in the input parameters for
the models, thus allowing us to obtain a robust model for 
the observed region. In particular the [\oii]/[\oiii] ratios are very sensitive
to the degree of ionization, and the [\oii], [\sii], and [\oi] to \hb\ 
line ratios give us information on the behaviour of the low
degree of ionization regions of the nebula. All these line ratios
are also sensitive to the density distribution adopted.

In sections~3.1 to~3.5 we discuss density bounded models
in all directions in agreement with the results presented in section~2. 
In section~3.6 we discuss two ionization bounded models which
provide independent support for the view that NGC~346 is a density bounded nebula.

\subsection{Constant density models}

The simplest models that can be computed are spherical with
constant density (models L.1, S.1, and B.1). The density is given by the
root mean square density, $N_{\rm e}$(rms), and takes the value 9.00~\cm3. This
value was derived from the \ha\ flux from Kennicutt \& Hodge (1986),
a $C$(\ha) of 0.10~dex (Paper I), a distance to the SMC of 64~kpc 
(Reid 1999, and references therein), and a radius of 210\arcsec.

Using these models we find that most of
the predicted line ratios differ considerably from the observed ones. In
particular the lines of low degree of ionization, such as those of
\oii\ and \sii, are very faint (see Table~2). 

By looking at the [\oii] and [\oiii] line
intensities it can be seen that the degree
of ionization increases in the B.1, A.1, S.1 sequence
this result permits us to estimate the differences in the models
due to the ionization field adopted.

\subsection{Models with filling factor different from 1.0}

Considering the filamentary aspect of NGC~346 (e.g. Ye et al. 1991),
the densities derived in Paper I,
and the weakness of the lines with low degree of ionization, we decided
to compute a model with a filling factor different from 1, and a 
constant $N_{\rm e}$(local).

The filling factor can be estimated by means of the relation:
\begin{equation}
    N_{\rm e}^2({\rm rms})=\epsilon N_{\rm e}^2({\rm local}),
\end{equation} 
where $N_{\rm e}$(rms) is the root mean square electron density, and 
$N_{\rm e}$(local) 
is the electron
density determined through a 
forbidden-line ratio or from the He~{\sc{i}} lines based on the maximum likelihood
method (see Paper I). The He~{\sc{i}} lines originate in the (He{~\sc{ii}}) zone,
therefore the density derived from them is called the (He{~\sc{ii}}) density.
 
We computed models of the L series with different filling factors and
the best fit to the observed \oii/\hb, \oiii/\hb, and \oiii/\oii\
ratios is provided by model L.4 with an $N_{\rm e}$(local) of 80\cm3\ and
a filling factor $\epsilon$ of 0.0127. By comparing the
line ratios predicted by model L.4 with observations (see Table~2) 
we find a much better fit than that provided by model L.1. Also in Table~2
we present models L.2 and L.3 to show the sensitivity of the fit to
the adopted density (or filling factor). 

We also computed models
of the S series with different filling factors and present them in Table~3.
Model S.2, which has the same density and filling factor as model
L.2, provides a considerably better fit to the observations
than the S.1 model, in agreement with the result obtained with the L series.
As with the case of the L.1 and S.1 models, the S.2 model shows a
higher degree of ionization than the L.2 model. By increasing the density
it is possible to obtain a better \oii/\hb\ fit with a model of the S series,
but the fit of the \oiii/\oii\ and the \oiii/\hb\ ratios becomes poorer.

Models with 0.005 $<$ $\epsilon$ $<$ 0.02 provide a much better fit to the
observations than models with $\epsilon$ = 1.00 (see Tables~2 and 3).

\subsection{Decreasing density models}

We have two determinations of $N_{\rm e}$(local), 
$N_{\rm e}$(He{~\sc{ii}})$ = 143 \pm 36$\cm3\
and $N_{\rm e}$(S~{\sc{ii}})$ = 50 \pm 15$\cm3. Since the sulfur emission
originates in the outer parts and the helium emission all over
the \hii\ region, we decided to compute two
models with higher density in the center and lower in the outer regions.

Model L.7 has the following decreasing density law:

\begin{equation}
N(r)=N_{0}\left(\frac{r}{r_{0}}\right) ^{-\alpha},
\end{equation}  
where $r_{0}$ and $N_{0}$ are, respectively, the inner radius and the density
at the inner radius.  We choose a density law with an
exponent $\alpha$=0.315, which gives a value of 180~\cm3\ at $r_{0} = 10^{19}$~cm
and a value of 70~\cm3\ at the boundary ($r = 2.0 \times 10^{20}$~cm). From the observational
constraint in the \ha\ flux we obtained that $\epsilon$ = 0.013.  

Model B.2 is made of two concentric spherical shells with $N_{\rm e}$(local) equal to 140 
and 50~\cm3\
respectively and the following radial intervals:  $10^{16}$ to $10^{20}$ and 
$10^{20}$ to $2.0 \times 10^{20}$~cm respectively.

Models L.7 and B.2 provide a better fit to the observations than models
L.1 and B.1, but do not provide a significantly better fit than L.4,
the main reference model.

\subsection{Models with different chemical composition}

Models L.5 and L.6 were computed with 0.1~dex higher and 0.1~dex lower heavy
element abundances than model L.4. The differences in the O/H value are
about 2$\sigma$ higher for model L.5 and 2$\sigma$ lower for model
L.6 than the values derived in Paper I.

Model L.5 produces a modest increase in respect to model L.4 in the $I(3727 + 5007)/I(H\beta)$
value without reaching the observed value. Alternatively the difference
between the predicted and the observed $I(4363)/I(5007)$ value becomes
larger than in model L.4.

Model L.6 produces a higher $I(4363)/I(5007)$ value than model L.4 without
reaching the observed value. Alternatively the difference between the
observed and predicted $I(3727 + 5007)/I(H\beta)$ value becomes larger
than in model L.4.

\subsection{Plane-parallel models}

We decided to study a completely different geometric
arrangement to see if it was possible to get a better
agreement between models and observations. In addition
to one dimensional models with spherical symmetry, it is
possible to compute one dimensional plane parallel models.
For these models we have adopted five observational restrictions
associated with the geometry: a) the distance to the object, b)
the subtended solid angle, c) the observed
\ha\ flux, d) an $N_{\rm e}$(local)=100~\cm3, and e) 
the incident flux per unit area
in the direction of the line of sight, that we
assume to be constant all over the face of the nebula.
The incident flux per unit area
is given by the ionizing stellar flux divided by the area
perpendicular to the line of sight.

The line intensities predicted by one of our plane parallel models (B.3)
are presented in Table~3. The model has an area
perpendicular to the line of sight of
$1.26 \times 10^{41}$~\cmdos\ and a length in the direction of the 
line of sight of $0.9 \times 10^{20}$~cm, which implies
an $\epsilon$ of 0.0216. The model provides a poorer fit than the spherical 
models because it shows a considerably lower degree of ionization than
the observations.

Other geometries require two or three dimensional photoionization
codes and are beyond the scope of this paper.

\subsection{Ionization bounded}

To explore further the possibility that NGC~346 is a density bounded
\hii\ region, we have studied three line ratios involving lines with low
degree of ionization: $I(\lambda6300)/I(H\beta$),
$I(\lambda6717)/I(H\beta$), and
$I(\lambda3727)/I(H\beta$). These ratios
are stronger in the outer zones of an ionization bounded \hii\ region than in 
the outer zones of a density bounded
\hii\ region. In Table~4 we show these line
ratios for each model as well as the observational values
from Paper I. For $I(\lambda6300)/I(H\beta$), all the density bounded models 
give
a value of the line ratio in agreement with the observational limit, and
models L.2 to L.6, B.2, and S.2 show good agreement with the
other two lines. 

In Table~4 we  present these three line ratios for two ionization 
bounded models,
IL.4 and IS.2. These models are similar to the density bounded models L.4 and 
S.2 the only
difference is that the intensity of the ionizing flux is 45\% smaller than in 
the density
bounded models so that all the ionizing photons are trapped by the nebula. For 
the IS.2
model the three line ratios are stronger than observed, while for the IL.4 model
the predicted $I(\lambda6300)/I(H\beta$) value is smaller than the upper limit
while the other two line ratios are stronger than observed. To reach agreement
between ionization bounded models and the observations it is necessary to reduce
$N_{\rm e}$(local) (and to increase the 
filling factor accordingly) entering into conflict with the observed 
$N_{\rm e}$(He{~\sc{ii}})$ = 143 \pm 36$~\cm3
and $N_{\rm e}$(S~{\sc{ii}})$ = 50 \pm 15$~\cm3\ values. Of the line ratios
presented in Table~4 the best discriminant between the density bounded 
and the 
ionization bounded models is the $I(\lambda6300)/I(H\beta$) value, therefore 
it is important to obtain a determination 
of this line ratio and not only an upper limit. The previous discussion supports
the result that NGC~346 is a density bounded nebula.

Models for NGC~346 with a covering factor of 0.55
will produce intermediate ionization
structures to those provided by density bounded and 
ionization bounded models. From the discussion of the
previous paragraph it follows that these models will 
produce better fits to the lines of low degree of ionization
than the ionization bounded models but poorer fits
than the density bounded models.

\section{Ionization Correction Factors and Abundances}

For most of the elements we observe only one or two stages of
ionization, therefore to obtain the total abundances relative to H
we need to determine the fraction of a given element present in the
unobserved stages of ionization. Therefore to obtain the total 
abundances we have made use of the following equations:
\begin{equation}
 \frac{N(\rm O)}{N(\rm H)} =
             \frac{N({\rm O^+})+N(\rm O^{++})}{N(\rm H^+)},
\end{equation} 
\begin{equation}           
 \frac{N(\rm N)}{N(\rm H)}  =  ICF({\rm N})~ N({\rm N}^{+})/N({\rm H}^{+}),
\end{equation} 
\begin{equation}           
 \frac{N(\rm Ne)}{N(\rm H)}  =  ICF({\rm Ne})~ N({\rm Ne}^{++})/N({\rm H}^{+}),
\end{equation} 
\begin{equation}
 \frac{N(\rm S)}{N(\rm H)}  =  ICF({\rm S})~ N({\rm S}^{+} + {\rm S}^{++})/N({\rm H}^{+}),
\end{equation} 
\begin{equation}
 \frac{N(\rm Ar)}{N(\rm H)}  =  ICF({\rm Ar}) 
             \frac{N({\rm Ar^{++}})+N(\rm Ar^{3+})}{N(\rm H^+)},
\end{equation}
\begin{equation}           
 \frac{N(\rm Fe)}{N(\rm H)}  =  ICF({\rm Fe})~ N({\rm Fe}^{++})/N({\rm H}^{+}); 
\end{equation}
where the ionization correction factors, $ICF's$, were estimated from the models
presented in this paper, see Table~5, and the ionic abundance ratios were taken
from Paper I.  To obtain the $N$(Fe$^{++}$)/$N$(H$^{+}$) value we made use of the
observed intensity of $\lambda$ 4568 in region A and its predicted intensity by the
models.  The total abundances are presented in Table~6.

The He/H value is given by (e.g. Sauer \& Jedamzik 2001; Peimbert 2001 and references
therein):

\begin{eqnarray}
\frac{N ({\rm He})}{N ({\rm H})} & = &
\frac {\int{N_e N({\rm He}^0) dV} + \int{N_e N({\rm He}^+) dV} + 
\int{N_e N({\rm He}^{++})dV}}
{\int{N_e N({\rm H}^0) dV} + \int{N_e N({\rm H}^+) dV}},
                                                \nonumber \\
& = & ICF({\rm He})
\frac {\int{N_e N({\rm He}^+) dV} + \int{N_e N({\rm He}^{++}) dV}}
{\int{N_e N({\rm H}^+) dV}}
\label{eICF}
,\end{eqnarray}

where $ICF({\rm He})$ is the helium ionization correction factor.
To determine a very accurate He/H ratio a very precise value of
$ICF({\rm He})$ is needed. As expected for our density 
bounded models it is found that $ICF({\rm He})$ = 1.000, because there is no
transition zone from ionized to neutral He and H regions.

\section{Discussion and Conclusions}
 
We have presented a set of photoionization models for NGC~346. The advantage
of these models relative to those models present in the literature of giant 
extragalactic \hii\ regions
is that we know the spectra of most of the ionizing stars and the
absolute V magnitude for all of them, therefore we did not have to assume
the IMF, the age of the cluster, the star formation rate, nor
the fraction of ionizing photons that are trapped by the nebula.

The models were computed with different geometries,
different density distributions, different chemical compositions
and different radiation fields. We did not compute models composed via 
a superposition of separate
individual Str\"omgren spheres because a considerable amount
of the ionizing flux escaped from the nebula implying that
the Str\"omgren spheres produced by each star overlap, and
that the single \hii\ region model is a good approximation.

By comparing the models with the observed 
line intensity ratios we obtain the following results:

1) Spherical models with constant density, with $\epsilon = 1.00$ 
(models L.1, S.1, and B.1), attain a high degree of ionization that is not 
observed. Moreover, these models have an $N_{\rm e}$(rms) = 9.00~\cm3, this density 
is smaller than those derived from the [\sii] and \hei\  
lines (see Paper I).

2) Spherical models (L.4, L.7, S.2 and B.2) 
can reproduce the observed
degree of ionization and show good agreement with the [\sii] and \heii\
densities.

3) Models L.7 and B.2 include more realistic density distributions, but their
predictions are not very different to those of models L.4 and S.2.

4) The $I$(\heii, 4686)/$I$(\hb) ratio is very small and is very
sensitive to small changes in the radiation field and the geometry of 
the \hii\ region. Model L.1 has the same geometry as models S.1 
and B.1 but shows
a difference of about an order of magnitude in the $I$(\heii,
4686)/$I$(\hb)
ratio. The differences are due to the three radiation fields
used which vary in the fraction of photons with an energy higher
than four Rydbergs, while in
model B.1 we use
a blackbody spectrum as an ionizing source, in models L.1 and S.1 we use two
different sets of stellar atmopheres (see Figure~1).

Models L.2 to L.6 have similar $I$(\heii, 4686)/$I$(\hb)
ratios while model L.7
shows a ratio that is about two orders of magnitude fainter than the
other L models. All the models of the L series have the same radiation
field, in this case the difference is due to the geometry, the L.7 model has
an inner hole of $10^{19}$~cm while the other L models have an
inner hole of $10^{16}$~cm, the different geometry dilutes the radiation 
field in the inner regions of the nebula causing the drop in the 
helium ionization degree. A similar drop in the ionization degree,
due to geometrical effects, is 
present in the plane parallel model B.3, which shows an
$I$(\heii, 4686)/$I$(\hb) ratio two orders of magnitude fainter than the
spherical models B.1 and B.2.

5) Plane parallel models present a considerably lower degree of ionization
than is observed in NGC~346. For example
they underestimate the $I$([\ariv], 4740)/$I$([\ariii], 7135) and the 
$I$([\oiii], 5007)/$I$([\oii], 3727) ratios. On the other hand 
spherical models (like L.4, L.7, S.2 and B.2) present line ratios
in good agreement with the observations.

6) None of the models are able to reproduce the observed $I$(3727 + 5007)/$I$(\hb)
ratio nor the $I$(5007)/$I$(4363) ratio. This is a crucial result which
implies that photoionization models predict lower temperatures than
those observed. This result probably implies that there are additional
heating sources not considered by the photoionization models.
Similar results were obtained by Stasi\'nska \& Schaerer (1999) for
I Zw 18, Luridiana, Peimbert, \& Leitherer (1999) for NGC 2363, and
Luridiana \& Peimbert (2001) for NGC 5461.

7) The total abundances of N, Ne, S, Ar, and Fe were obtained from
the $ICF's$ predicted by model L.4. By comparing the empirical $ICF's$ in Paper I
with those derived from model L.4 we find excellent agreement
for Ne and Ar, good agreement for S, and only a fair agreement for N. Based 
on empirical methods and on photoionization models it has been 
found that the O$^+$ zone coincides with the N$^+$ zone
and consequently that the N/O ratio is equal to the N$^+$/O$^+$ ratio
(e. g. Peimbert \& Costero 1969; Garnett 1990; Mathis \& Rosa 1991; Stasinska \&
Schaerer 1997, Paper I); 
alternatively the N abundance derived from the NGC~346 {\sc Cloudy} models 
differs by about a
factor of two from that derived from the N$^+$/O$^+$ ratio (see Table~6),
Stasi\'nska \& Schaerer (1997) have studied this discrepancy and conclude that it
is due to the different ionizing spectra used by different authors.

From a detailed comparison between stellar ionizing photon input and 
the observed \ha\ flux we infer that some 45\% of the photons produced
by the ionizing stars escape from NGC~346. This result is independent
of the geometry and of the density distribution of the nebula, and
implies that this object must be a major source of ionizing flux 
for the surrounding diffuse interstellar medium. The ionization bounded
models we have tried fail completely to match the intensities of
lines with a low degree of ionization, but these can be matched using
density bounded models, which supports strongly the conclusion that
NGC~346 is density bounded.

It is a pleasure to acknowledge fruitful discussions with Valentina Luridiana and
Antonio Peimbert. We are also very grateful to Grazyna Stasi\'nska for a thorough
reading of a previous version of this paper and many excellent suggestions.
M.R. acknowledges financial help from DGES grant PB97-0219 which
facilitated her stay at the UNAM during a major phase of this study.

\clearpage

\begin{deluxetable}{lrcrlrl}
\tablecaption{Spectral Types, Temperatures, Mass, and Ionizing Fluxes
\label{spectraltypes}}
\tablewidth{0pt}
\tablehead{
\colhead{Spectral Type} & \colhead{$T$} & \colhead{$M$} &
\multicolumn{2}{c}{Number of stars} & \multicolumn{2}{c}{Ionizing Flux}\\
& \colhead{(K)} & \colhead{(\Msun)} &&& \multicolumn{2}{c}{($10^{39}$erg s$^{-1}$)}}
\startdata
O3V        & 51230\tablenotemark{a} & 51.3\tablenotemark{a} &  1 &      &   2.02 &        \\
O4V        & 48670\tablenotemark{a} & 44.2\tablenotemark{a} &  1 &      &   1.43 &        \\
O5.5V      & 44840\tablenotemark{a} & 35.5\tablenotemark{a} &  3 &  (1) &   2.46 & (0.82) \\
O6V        & 43560\tablenotemark{a} & 33.1\tablenotemark{a} &  3 &  (1) &   2.07 & (0.69) \\
O6.5V      & 42280\tablenotemark{a} & 30.8\tablenotemark{a} &  2 &  (1) &   1.14 & (0.57) \\
O7V        & 41010\tablenotemark{a} & 28.8\tablenotemark{a} &  3 &  (2) &   1.40 & (0.94) \\
O7.5V      & 39730\tablenotemark{a} & 26.9\tablenotemark{a} &  2 &  (2) &   0.76 & (0.76) \\
O8V        & 38450\tablenotemark{a} & 25.1\tablenotemark{a} &  6 &  (7) &   1.88 & (2.20) \\
O8.5V      & 37170\tablenotemark{a} & 23.6\tablenotemark{a} &  1 & (11) &   0.26 & (2.82) \\
O9V        & 35900\tablenotemark{a} & 22.1\tablenotemark{a} &  2 &      &   0.43 &        \\
O9.5V      & 34620\tablenotemark{a} & 20.8\tablenotemark{a} &  5 &      &   0.86 &        \\
O5.5I      & 43210\tablenotemark{a} & 45.4\tablenotemark{a} &  1 &      &   4.93 &        \\
O7I        & 38720\tablenotemark{a} & 37.4\tablenotemark{a} &  1 &      &   1.77 &        \\
WN4(5980A) & 52000\tablenotemark{b} & 18.0\tablenotemark{c} &  1 &      &   7.01 &        \\
O7Ia(5980B)& 31135\tablenotemark{b} & 37.4\tablenotemark{a} &  1 &      &   2.86 &        \\
\\
Total        &  \nodata & \nodata & 33 & (25) &  31.28 & (8.80)
\enddata
\tablenotetext{a}{Vacca et al. (1996).}
\tablenotetext{b}{Schweickhardt \& Schmutz(1999).}
\tablenotetext{c}{Koenigsberger et al. (1998)}
\end{deluxetable}

\clearpage

\begin{deluxetable}{lrrrrrrrrr}
\tablecaption{Photoionization models based on a set of atmospheres with $Z_{*}=0.2Z_{\odot}$ 
\label{ModelA}}
\rotate
\tabletypesize{\scriptsize}
\tablewidth{0pt}
\tablehead{
\colhead{Line ratios\tablenotemark{a}} & \colhead{Reg. A} & \colhead{Reg 3
  and 13} & \colhead{Mod L.1} & \colhead{Mod L.2} & \colhead{Mod L.3} &  \colhead{Mod L.4} 
  & \colhead{Mod L.5} &\colhead{Mod L.6} &\colhead{Mod L.7}} 
\startdata
$[$\oiii$]~I(\lambda5007)/$[\oii]$~I(\lambda$3727) &  0.738 &  0.750 & 1.633 &
 0.492 & 0.383 & 0.595 &0.597 & 0.578 & 0.769 \\
$[$\siii$]~I(\lambda6312)/$[\sii]$~I(\lambda$6725) & -0.857 & -0.744 & -0.004 &
 -0.883 & -0.974 & -0.802 &  -0.847 &  -0.776 & -0.659 \\
$[$\ariv$]~I(\lambda4740)/$[\ariii]$~I(\lambda$7135) & -1.111 &-1.392 & -0.614 &
 -1.233 &-1.292 & -1.179 & -1.214  &  -1.155  & -1.136\\
$[$\oiii$]~I(\lambda4363)/I(\lambda$5007) & -1.854 &-1.915 & -2.041 &
 -2.015 &-2.010 & -2.021 & -2.100  &  -1.951  & -2.018\\
\heii$~I(\lambda4686)/$\hei$~I(\lambda$4471) & -1.155 &-1.430 & -0.905 &
 -1.286 &-1.337& -1.242 &  -1.243  &  -1.243 & -3.209\\
$[$\sii$]~I(\lambda6716)/I(\lambda$6731) &  0.144 & 0.142 & 0.152 &
  0.116 & 0.106 & 0.124 & 0.124  &  0.123 & 0.123\\
$[$\sii$]~I(\lambda4069+4076)/I(\lambda$6725) & -1.115 &-0.940 & -1.104 &
 -1.050 &-1.038&-1.060 & -1.086  &  -1.037 & -1.060\\
$[$\oiii$]~I(\lambda5007)/I(H\beta$) & 0.735  & 0.710 & 0.625 &
 0.541 &0.518& 0.559 &  0.579    &   0.526  & 0.609 \\
$[$\oii$]~I(\lambda3727)/I(H\beta$) &  -0.003 &-0.040 & -1.008 &
 0.049 &0.135& -0.036 &  -0.018   &   -0.052 & -0.160\\
\heii$~I(\lambda4686)/I(H\beta$) & -2.571 &-2.844 & -2.419 &
 -2.804 &-2.856& -2.759 & -2.753  & -2.766 & -4.708\\
\\
\\

$L$(\ha)(dex)  & \nodata & \nodata & 38.987 & 38.978 & 38.987 & 38.977 &
38.988 & 38.981 & 38.989\\
$N_e$(rms)(\cm3)& \nodata & \nodata & 9.00\phd\phd &  9.00\phd\phd   &
9.00\phd\phd & 9.00\phd\phd & 9.00\phd\phd & 9.00\phd\phd & 9.00\phd\phd \\ 
radius ($\mbox {10}^{\mbox {20}}$ cm)   & \nodata & \nodata & 2.00\phd\phd & 2.01\phd\phd &
2.01\phd\phd   &  2.01\phd\phd & 2.00 \phd\phd& 2.04 \phd\phd& 2.02\phd\phd  \\ 
$N_e$(local)    & \nodata& \nodata & 9.00\phd\phd &  100 & 130 & 80 & 80 & 80 
& \nodata \\
$\epsilon$              & \nodata & \nodata & 1.00\phd\phd & 0.0081 &  0.005 &
0.0127 & 0.0127 & 0.0127 & 0.013 \\
$[{\mbox O}]$\tablenotemark{b}          & \nodata
  & \nodata  & 8.11\phd\phd &
8.11\phd\phd & 8.11\phd\phd & 8.11\phd\phd & 8.21\phd\phd& 8.01\phd\phd& 8.11\phd\phd\\

$T_e$(rad)K    &\nodata  & \nodata & 11200 &  11400 & 11500 & 11400 & 10700 & 
12000 & 11400\\
$T_e$(vol)K    & \nodata & \nodata & 10800 &  11200 & 11300 & 11100 & 10600 & 
11700 & 11100\\ 
$T_e$([\oiii])K &  13070 &   12430 & 10800 &  11300 & 11300 & 11200 & 10600 & 
11800 & 11200 \\
\enddata
\tablenotetext{a} {Given by log $I(\lambda_1)/I(\lambda_2$).}
\tablenotetext{b} {Gaseous abundances given by 12 + log $N$(O)/$N$(H).}
\end{deluxetable}
\clearpage

\begin{deluxetable}{lrrrrrrr}
\tablecaption{Photoionization models based on other sets of stellar atmospheres\label{ModelSB}}
\rotate
\tabletypesize{\scriptsize}
\tablewidth{0pt}
\tablehead{
\colhead{Line ratios\tablenotemark{a}} & \colhead{Reg. A} & \colhead{Reg 3
  and 13} & \colhead{Mod S.1\tablenotemark{b}} & \colhead{Mod S.2
  \tablenotemark{b}} & \colhead{Mod B.1\tablenotemark{c}} & \colhead{Mod B.2\tablenotemark{c}} 
  & \colhead{Mod B.3\tablenotemark{c}}} 
\startdata
$[$\oiii$]~I(\lambda5007)/$[\oii]$~I(\lambda$3727) &  0.738 &  0.750 & 1.918 &0.716 & 1.435 &
0.759 &   0.190\\
$[$\siii$]~I(\lambda6312)/$[\sii]$~I(\lambda$6725) & -0.857 & -0.744 & -0.065 &-0.927 & -0.054 & 
-0.587 & -0.882\\
$[$\ariv$]~I(\lambda4740)/$[\ariii]$~I(\lambda$7135) & -1.111 &-1.392 & -0.346 &-0.994 & -0.812 &
-1.156 & -2.581\\
$[$\oiii$]~I(\lambda4363)/I(\lambda$5007) & -1.854 &-1.915 &  -1.981 &-1.976 & -2.077 &
-2.039 & -2.078\\
\heii$~I(\lambda4686)/$\hei$~I(\lambda$4471) & -1.155 &-1.430 &  0.293 &0.183 & 0.359 &
0.359 & -1.916\\
$[$\sii$]~I(\lambda6716)/I(\lambda$6731) &  0.144 & 0.142 & 0.150 &0.117 & 0.152&
0.112 & 0.116\\
$[$\sii$]~I(\lambda4069+4076)/I(\lambda$6725) & -1.115 &-0.940 & -1.080 &-1.043 & 
-1.125 & -1.065  & -1.065 \\
$[$\oiii$]~I(\lambda5007)/I(H\beta$) & 0.735  & 0.710 & 0.663 &0.607 & 0.595&
   0.585 & 0.478\\
$[$\oii$]~I(\lambda3727)/I(H\beta$) &  -0.003 &-0.040 & -1.255 &-0.109 &
-0.840 & -0.174 & 0.288  \\
\heii$~I(\lambda4686)/I(H\beta$) & -2.571 &-2.844 & -1.255  &-1.360 &
-1.132 & -1.137 & -3.338 \\
\\
\\

$L$(\ha)(dex)  & \nodata & \nodata & 38.981 &38.973 & 38.975 & 38.989 & 38.993\\
$N_e$(rms)(\cm3)& \nodata & \nodata & 9.00\phd\phd &  9.00\phd\phd   &
9.00\phd\phd  &  9.00\phd\phd &  9.00\phd\phd \\ 
radius ($\mbox {10}^{\mbox {20}}$ cm)   & \nodata & \nodata & 2.01 & 2.01 &
1.94 & 2.00 & \nodata  \\ 
$\mbox {$R$}_{\mbox 1}$ ($\mbox {10}^{\mbox {20}}$ cm)  & \nodata & \nodata &
\nodata & \nodata & \nodata & 1.01 & \nodata\\
$N_e$(local)    & \nodata& \nodata & 9.00 & 100 & 9.00 & (140,50) & 100 \\
$\epsilon$              & \nodata & \nodata & 1.00 & 0.0081 & 1.00 &
0.0175 & 0.0216\\
$[{\mbox O}]$\tablenotemark{d}  & \nodata &    \nodata     & 8.11 & 8.11 & 8.11 & 8.11 & 8.11\\

$T_e$(rad)K    &\nodata  & \nodata & 11700 & 11700 & 10800 & 11100 & 10900\\
$T_e$(vol)K    & \nodata & \nodata & 11200  &11400 & 10400 & 10800 & 10900\\ 
$T_e$([\oiii])K &  13070 &   12430 & 11300  &11500 & 10400 & 10800 & 11000\\
\enddata
\tablenotetext{a} {In units of log $I(\lambda_1)/I(\lambda_2$).}
\tablenotetext{b} {$Z_{*}=Z_{\odot}$}
\tablenotetext{c} {Blackbody ionizing spectrum.}
\tablenotetext{d} {Gaseous abundances given by 12 + log $N$(O)/$N$(H).}
\end{deluxetable}
\clearpage

\begin{deluxetable}{cccc}
\tablecaption{Line ratios of [O~{\sc{i}}], [S~{\sc{ii}}], and [O~{\sc{ii}}]
\label{intoi}}
\tablewidth{0pt}
\tablehead{
\colhead{Models} & \colhead{$I(\lambda6300)/I(H\beta$)} &
\colhead{$I(\lambda6717)/I(H\beta$)} & \colhead{$I(\lambda3727)/I(H\beta$)}} 
\startdata
$Z_{*}=0.2Z_{\odot}$ & & & \\
L.1 & -5.402 & -2.244 & -1.008 \\ 
L.2 & -3.155 & -1.140 &  0.049 \\
L.3 & -2.953 & -1.042 &  0.135 \\ 
L.4 & -3.340 & -1.230 & -0.036 \\ 
L.5 & -3.289 & -1.279 & -0.018 \\ 
L.6 & -3.362 & -1.174 & -0.052 \\ 
L.7 & -3.609 & -1.327 & -0.334 \\ 
Blackbody Spectrum & & &\\
B.1 & -5.277 & -2.136 & -0.840 \\
B.2 & -3.878 & -1.433 & -0.174 \\ 
B.3 & -3.328 & -1.029 &  0.288 \\ 
$Z_{*}=Z_{\odot}$ & & &\\
S.1 & -5.600  & -2.297 & -1.255 \\ 
S.2 & -3.219  & -1.116 & -0.109 \\
IL.4 \tablenotemark{a} & -2.270 &  -0.861 & 0.234 \\
IS.2 \tablenotemark{a} & -1.707 &  -0.844 & 0.193 \\
Observations\tablenotemark{b} & $<$ -2.016 &  -1.284 & -0.040 \\

\enddata
\tablenotetext{a} {Ionization bounded models, see text}
\tablenotetext{b} {For Region A and from Paper I}
\end{deluxetable}
\clearpage

\begin{deluxetable}{ccc}

\tablecaption{Ionization Correction Factors for Region A
\label{Table X}}
\tablewidth{0pt}
\tablehead{
\colhead{Element} & \colhead{Paper I} & \colhead{This paper}}
\startdata
N  &  5.62  &  11.22 \\
Ne &  1.23  &   1.29 \\
S  &  1.98  &   1.29 \\
Ar &  1.21  &   1.22   \\
Fe & \nodata&    8.00:  \\

\enddata
\end{deluxetable}
\clearpage

\begin{deluxetable}{cccccc}
\tablecaption{Total Abundances\tablenotemark{a}
\label{Table X+1}}
\tablewidth{0pt}
\tablehead{
\colhead{Element} & \colhead{NGC~346} & \colhead{NGC~346} & \colhead{Sun\tablenotemark{b}} & \colhead{Ori\'on
  \tablenotemark{c}} & \colhead{M17\tablenotemark{d}} \\
& \colhead{Paper I} & \colhead{This paper} & & & }
\startdata
N      &    6.51    &    6.81    &    7.92    &  7.78    &   7.90 \\     
O\tablenotemark{e}   &    8.15    &    8.15    &    8.83    &  8.72    &   8.87 \\
Ne     &    7.30    &    7.32    &    8.08    &  7.89    &   8.02 \\
S      &    6.59    &    6.40    &    7.33    &  7.17    &   7.31 \\
Ar     &    5.82    &    5.82    &    6.40    &  6.49    &   6.60 \\
Fe\tablenotemark{f}  &   \nodata   &    6.16    &    7.50    &  6.11    &   6.69 \\
\enddata
\tablenotetext{a} {In units of 12 + Log $N$(X)/$N$(H).}
\tablenotetext{b} {Grevesse \& Sauval (1998).}
\tablenotetext{c} {Esteban et al. (1998).}
\tablenotetext{d} {Peimbert, Torres-Peimbert, \& Ruiz (1992); Esteban et al. (1999).}
\tablenotetext{e} {Gas plus dust for the \hii\ regions.}
\tablenotetext{f} {Gaseous content only.}

\end{deluxetable}

\begin{figure}
\plotone{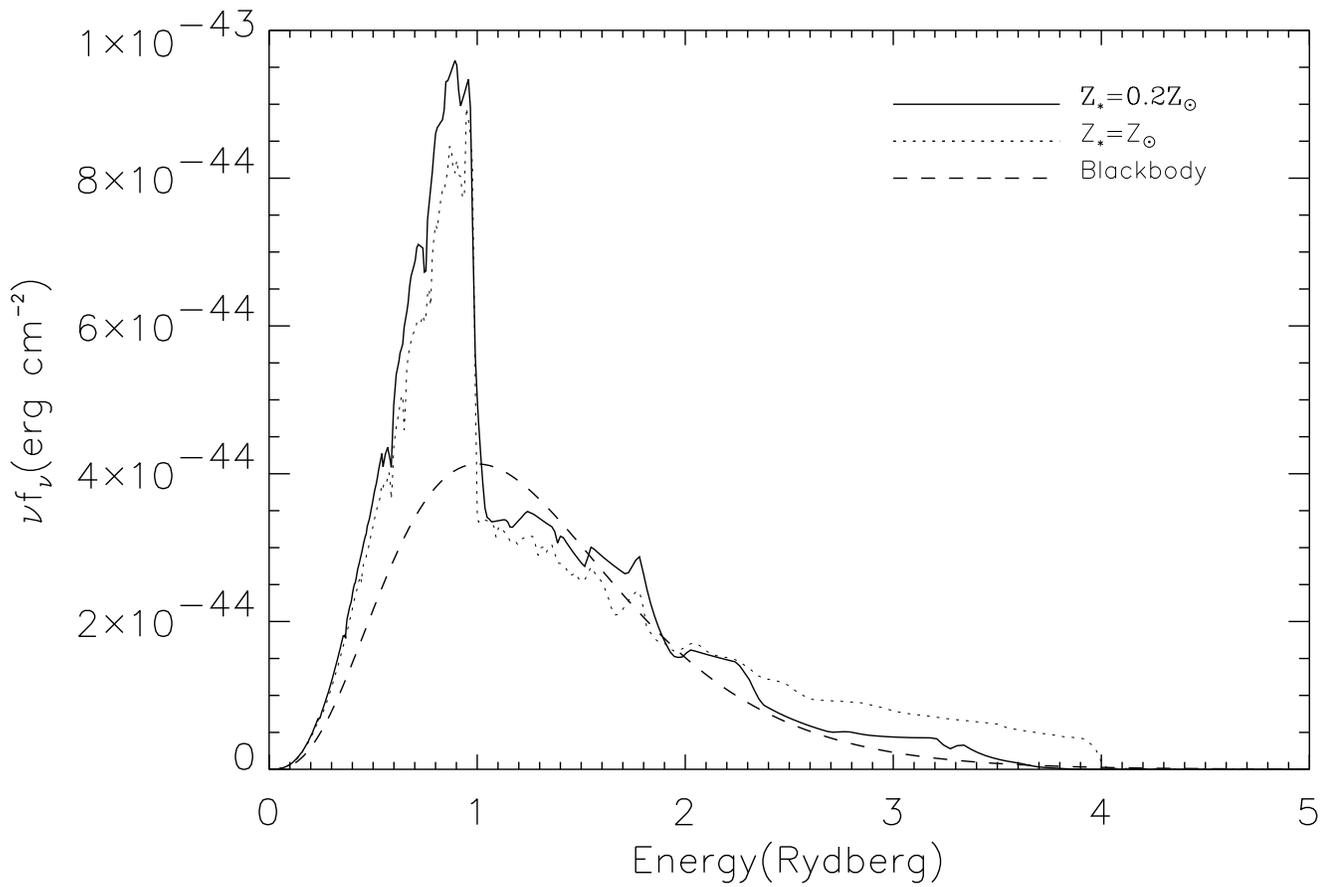}
\label{cont}
\caption{Ionizing continuum spectra used in the models. All the spectra are
  normalized to the same total number of ionizing photons.}
\end{figure}

\begin{figure}
\plotone{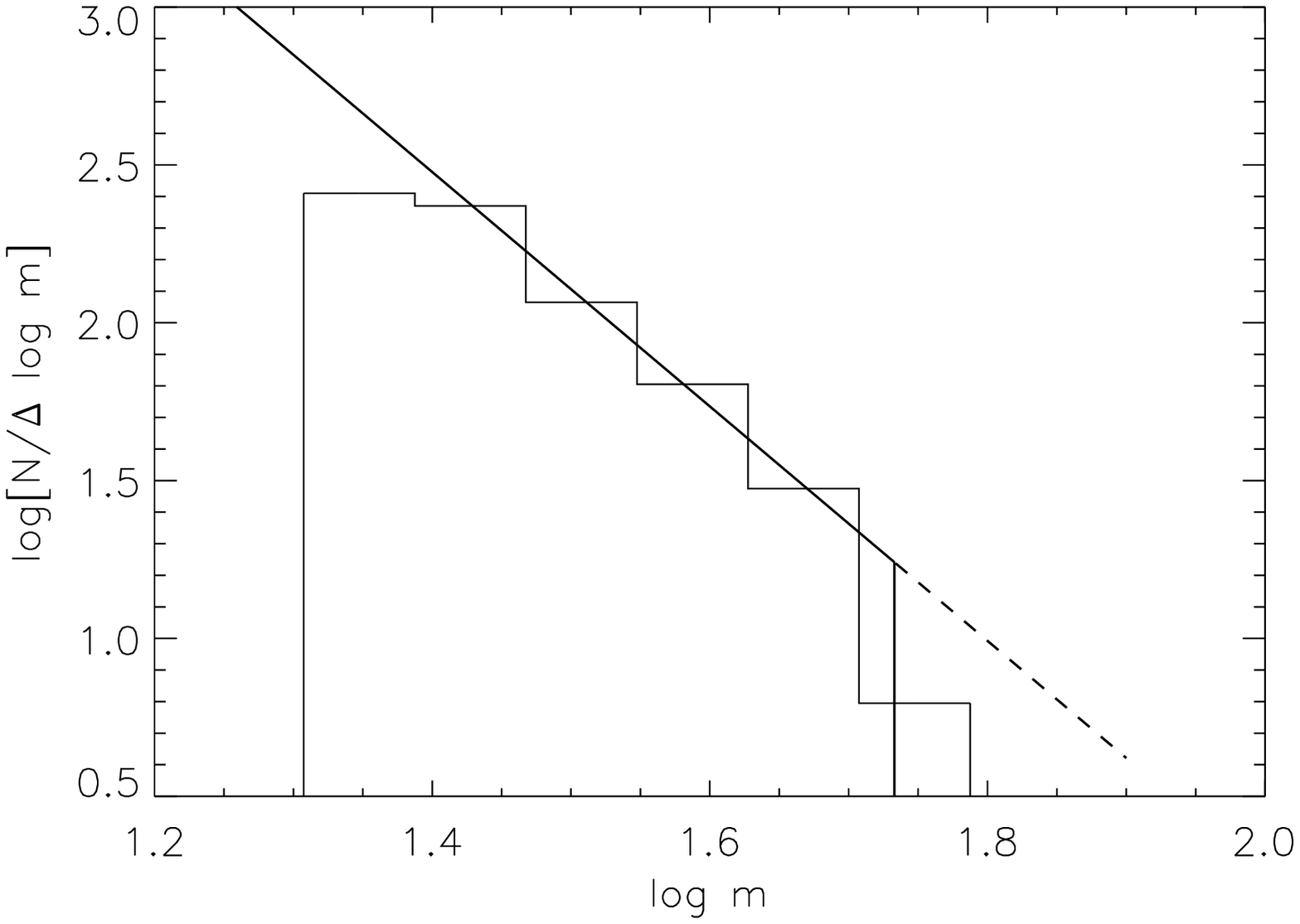}
\label{fmi}
\caption{IMF derived from Table~1. The straight line represents the
IMF derived from the four central bins and it has a slope of
-3.7$\pm$0.4. This IMF has an upper mass cutoff of 54.1~\Msun\ and it is
only representative for masses higher than 24~\Msun.}
\end{figure}


\bigskip

\begin{thebibliography}{}

\bibitem{}
Cervi\~no, M., Luridiana, V., \& Castander, F.J. 2000, 
\aap, {360}, L5

\bibitem{}
Esteban, C., Peimbert, M.,
Torres-Peimbert, S., \& Escalante, V.  1998, \mnras, 295, 401

\bibitem{}
Esteban, C., Peimbert, M.,
Torres-Peimbert, S.,\& Garc\'{\i}a-Rojas,  J. 1999, RevMexAA, 35, 85

\bibitem{}
Ferland, G. J. 1996, Hazy, a Brief Introduction
to {\sc Cloudy 90} (Univ. Kentucky Dept. Phys. Astron. Internal Rep.)

\bibitem{}
Ferland, G. J., Korista, K. T., Verner,
D. A., Ferguson, J. W., Kingdon, J. B., \& Verner, E. M. 1998, \pasp, 110, 761

\bibitem{}
Garnett, D. R. 1990, \apj, 363, 142

\bibitem{}
Garnett, D. R., Skillman, E. D., Dufour, R. J.,
Peimbert, M., Torres-Peimbert, S., Terlevich, R. J., Terlevich,
E., \& Shields, G. A., 1995, \apj, 443, 64

\bibitem{}
Grevesse, N. \& Sauval, A. J. 1998,
Space Sci. Rev., 85, 161

\bibitem{} 
Humphreys, R. M., \& McElroy, D. B. 1984, ApJ, 284, 565

\bibitem{}
Kennicutt, R. C., Jr., \& Hodge,
P. W. 1986, \apj, 306, 130

\bibitem{}
Koenigsberger, G., Auer, L. H., Georgiev, L., \& Guinan, E. 1998, \apj, 496, 934
\bibitem{}
Kurucz, R. L., 1991, in Proceedings of the
Workshop on Precision Photometry: Astrophysics of the Galaxy,
A. C. Davis Philip, A. R. Upgren, \& K. A. James (Davis, Schenectady), p 27 

\bibitem{}
Lejeune, Th., Cuisinier, F., \& Buser,
R. 1997, 125, 229

\bibitem{}
Luridiana, V., \& Peimbert, M.
2001, \apj, 553, 663

\bibitem{}
Luridiana, V., Peimbert, M., \& Leitherer, C. 1999, \apj,
527, 110

\bibitem{}
Massey, P., Parker, J. W., \& Garmany, C. D. 1989, \aj, 98, 1305

\bibitem{}
Mathis, J. S., \& Rosa, M. R. 1991, \aap, 245, 625

\bibitem{}
Mihalas, D. 1972, Non-LTE Model Atmosphere
for B\&O Stars (NCAR-TN/-STR-76; Denver:NCAR)

\bibitem{}
Oey, M. S., \& Kennicutt, R. C., Jr. 1997, \mnras, 291, 827

\bibitem{} Peimbert, M. {\it Revista Mexicana de 
Astronom\'{\i}a y Astrof\'{\i}sica, Serie de Conferencias}, 
in press (astro-ph/0106063), 2001.

\bibitem{} 
Peimbert, M. \& Costero, R. 1969, Bol. Obs. Tonantzintla y Tacubaya, 5, 3

\bibitem{}
Peimbert, M., Peimbert, A., \& Ruiz, M.T. 2000,
\apj, 541, 688, Paper I

\bibitem{} Peimbert, M., Torres-Peimbert, S., \& Ruiz, M.T. 1992,
RevMexAA, 24, 155

\bibitem{} Reid, I. N. 1999, \araa, 37, 191

\bibitem{} Sauer, D., \& Jedamzik, K. 2001,\aap, submitted (astro-ph/0104392)

\bibitem{} Schaerer, D., \& de Koter,
  A. 1997, \aap, 322, 598

\bibitem{}
Schweickhardt, J., 
\& Schmutz, W. 1999, in IAU Symp. 193, Wolf-Rayet Phenomena in Massive Stars
and Starburst Galaxies, ed. K. A. ven der Hucht, G. Koenigsberger \&
P. R. J. Enens (San Francisco: ASP), 101

\bibitem{} Stasi\'nska, G., \& Schaerer, D. 1997, \aap, 322, 615

\bibitem{}
Stasi\'nska, G., \& Schaerer, D. 1999, \aap, 351, 72

\bibitem{}
Vacca, W.D., Garmany, C. D., \& Shull, J. 
M. 1996, \apj, 460, 914

\bibitem{}
Ye, T., Turtle, A.J., \& Kennicutt, R. C., Jr. 1991,
\mnras, 249, 722

\bibitem{}
Zurita, A., Rozas, M., \& Beckman, 
J. E. 2000, \aap, 363, 9

\end{thebibliography}
\end{document}